\documentclass[twocolumn]{aastex7}

\newcommand{\jjj}{HESS J1813-126~}
\newcommand{\swift}{{\it Swift}-XRT~}
\newcommand{\psr}{PSR J1813-1246~}

\begin{document}

\title{Constraints on diffuse X-ray Emission from the TeV halo Candidate HESS J1813-126}

\correspondingauthor{David Guevel}
\email{guevel@wisc.edu}

\author[0000-0002-0870-2328]{David Guevel}
\affiliation{Department of Physics, Wisconsin IceCube Particle Astrophysics Center, University of Wisconsin, Madison, WI, USA}
\email{guevel@wisc.edu}

\author[0000-0001-5624-2613]{Kim L. Page}
\affiliation{School of Physics \& Astronomy, University of Leicester, 
University Road,  
Leicester, LE1 7RH, United Kingdom}
\email{klp5@leicester.ac.uk}

\author[0000-0002-9709-5389]{Kaya Mori}
\affiliation{Columbia Astrophysics Laboratory, Columbia University, 538 West 120th Street, New York, NY 10027, USA}
\email{kaya@astro.columbia.edu}

\author[0000-0002-7851-9756]{Amy Lien}
\affiliation{University of Tampa, 
401 W Kennedy Blvd,  
Tampa, FL, USA}
\email{yarleen@gmail.com}

\author[0000-0002-5387-8138]{Ke Fang}
\affiliation{Department of Physics, Wisconsin IceCube Particle Astrophysics Center, University of Wisconsin, Madison, WI, USA}
\email{kefang@physics.wisc.edu}

\begin{abstract}

Extended regions of very high energy $\gamma$-ray emission associated with middle-aged pulsars have been found by $\gamma$-ray observatories. 
These regions, called TeV halos or pulsar halos, are thought to be created when energetic electrons from a pulsar or pulsar wind nebula transport into interstellar medium and undergo inverse Compton scattering with the cosmic microwave background radiation.
The same electrons are expected to emit synchrotron emission in the X-ray band in the interstellar magnetic field.
\jjj is a pulsar halo candidate from which TeV $\gamma$-ray emission with extension 0.21\degr and a hard $E^{-2}$ spectrum is observed. 
We searched for the synchrotron component of this pulsar halo with {\it Swift}-XRT. In particular, we observed two fields within the region covered by \jjj  for 35 ksec each and a region nearby as a background reference for 10 ksec.
We find no evidence for excess X-ray emission from the two observations near \jjj and place an upper limit differential flux of $4.32\times 10^{-4}\, \rm  keV^{-1}\, cm^{-2}\,s^{-1} $ and $5.38\times 10^{-4}\, \rm  keV^{-1}\, cm^{-2}\,s^{-1} $ at 1 keV assuming an $E^{-2}$ power law spectrum. The non-detection implies that the magnetic field inside the halo is not significantly enhanced compared to the average  Galactic magnetic field.

\end{abstract}

\section{Introduction} \label{sec:intro}
Very high energy (0.1-100~TeV) $\gamma$-ray observatories have recently found numerous extended emission regions associated with middle-aged pulsars \citep{abdoMILAGROOBSERVATIONSMULTITeV2009, abeysekaraExtendedGammaraySources2017a, 3HWC, LHAASO:2021crt}. These extended emission regions, referred to as TeV halos or pulsar halos \citep{lindenUsingHAWCDiscover2017}, 
are produced when electrons are accelerated within the pulsar wind nebula (PWN) and diffuse out from the central region.
The acceleration and transport of these electrons are poorly understood, but the TeV $\gamma$-ray emission is likely produced by the inverse Compton (IC) scattering of relativistic electrons with the cosmic microwave background (CMB) \citep{2019PhRvD.100l3015D}.

Pulsar halos are known for their mysterious extension--the extensions of the observed TeV halos require the electron diffusion to be suppressed by several orders of magnitude relative to the interstellar medium \citep{sudohTeVHalosAre2019}. Recent observations by the HAWC Observatory of TeV halo emission around a population of middle-aged pulsars, but not millisecond pulsars \citep{PhysRevD.111.043014}, further suggest that pulsar halos may be a commonly existing phenomenon. These findings indicate that electron acceleration and confinement are related to the pulsar properties.

The origin of TeV halos remains poorly understood \citep{Lopez-Coto:2022igd}. Investigating the synchrotron component of pulsar halos provides an independent approach to understanding their formation mechanisms and probing the magnetic field strength within these structures. The synchrotron radiation of the electrons that produce the TeV halos should peak at $E_{\rm syn} =1.3\, (B / 3\,\mu {\rm G})(E_e / 100\,{\rm TeV})^2\,\rm keV$, where $B$ is the magnetic field strength and $E_e$ is the electron energy. This makes the X-ray band a promising window for identifying potential counterparts.
 


Numerous efforts have been made to detect counterparts to TeV halos.
Searches for X-ray counterparts have not revealed a conclusive counterpart in the Geminga pulsar as well as four additional pulsars \citep{manconiGemingasPulsarHalo2024, khokhriakovaSearchingXrayCounterparts2024}.
\citet{niuDetectionExtendedXray2025} found excess X-ray emission in the Monogem Ring with a very soft spectrum ($E^{-3.7}$) and a morphology that is much more compact than the TeV emitting region suggesting a spatially varying magnetic field.
Synchrotron radiation may also appear in radio wavelengths; \citet{hooperSearchingSynchrotronEmission2024} searched for radio emission from a TeV halo without a significant detection.

In this work,  we present constraints on X-ray emission from a  TeV halo candidate, HESS J1813-126. This source was first detected by the H.E.S.S. telescope \citep{HGPS} near the galactic anticenter a few degrees above the galactic plane. 
It has an $E^{-2}$ spectrum and an extended Gaussian morphology with a width of $0.21\degr$. This source is also detected by the HAWC Observatory as 3HWC~J1813-125  with a spectral index of 2.81 at 7~TeV \citep{3HWC}. 
The source was also detected by the LHAASO KM2A and WCDA with $E^{-3.66}$ and $E^{-2.26}$ spectra respectively \citep{caoFirstLHAASOCatalog2024a}.
The only likely associated source is a pulsar called PSR J1813-1246.
PSR J1813-1246 was discovered shortly after the launch of {\it Fermi} in a search of pulsation data \citep{abdoDetection16GammaRay2009}.
The period and period derivative of \psr are 48.1 ms and $1.76\times 10^{-14}$ s s$^{-1}$ suggesting a spindown luminosity of $6.24 \times 10^{36}$ erg s$^{-1}$ and characteristic age of 43~kyr \citep{abdoSECONDFERMILARGE2013a}.
No radio emission was observed from \psr by a dedicated Green Bank Observatory measurement \citep{abdoSECONDFERMILARGE2013a}.
No extended pulsar wind nebula was found in a search of X-ray data from Chandra and XMM-Newton by \citet{marelliPuzzlingHighenergyPulsations2014}.
Because it is located near the galactic anticenter, the background galactic X-ray emission is lower than  the galactic ridge or inner galaxy regions.
This offers a unique opportunity to search for faint diffuse emission from a TeV halo.

Below we  describe the \swift observations and data processing in Section 2. We explain the spectral fitting process in Section 3 and discuss the physical implications in Section 4.

\section{Swift Observations} \label{sec:swift}
\subsection{Observations}
HESS J1813-126 is an extended Gaussian source with a radius of 0.21\degr.
A conventional X-ray analysis uses a nearby region within the same exposure to estimate the background.
When a diffuse source fills the entire field of view, the background estimation must be done with one or more additional observations far enough from the source to avoid the diffuse emission.
We observed \jjj with \swift at multiple points to sample the X-ray emission from the halo over a range of radii.
Analysis of very diffuse objects have a very high background because the astrophysical and instrumental background are proportional to the detector area.
\swift was chosen because it has a low instrument background and wide field of view.
The field of view \swift has a 23.6\arcmin\, diameter, so it can just barely capture the bulk of the TeV halo in one observation if the center of the halo is exactly at the center of the \swift field of view.
We observed two targets overlapping the TeV emitting region.
The first was offset from the halo center by 15\arcmin, the second was offset by 30\arcmin.
These are referred to as HESS 1 and HESS 2.
We also performed a background observation well outside the 20\arcmin\, extent of the TeV halo.
{\it Swift} observed the on-source fields for 35 and 32 ksec and the background field for 9 ksec.
The background observation is referred to as HESS 3.

\begin{figure}[t!]
\centering
   \includegraphics[width=0.49\textwidth]{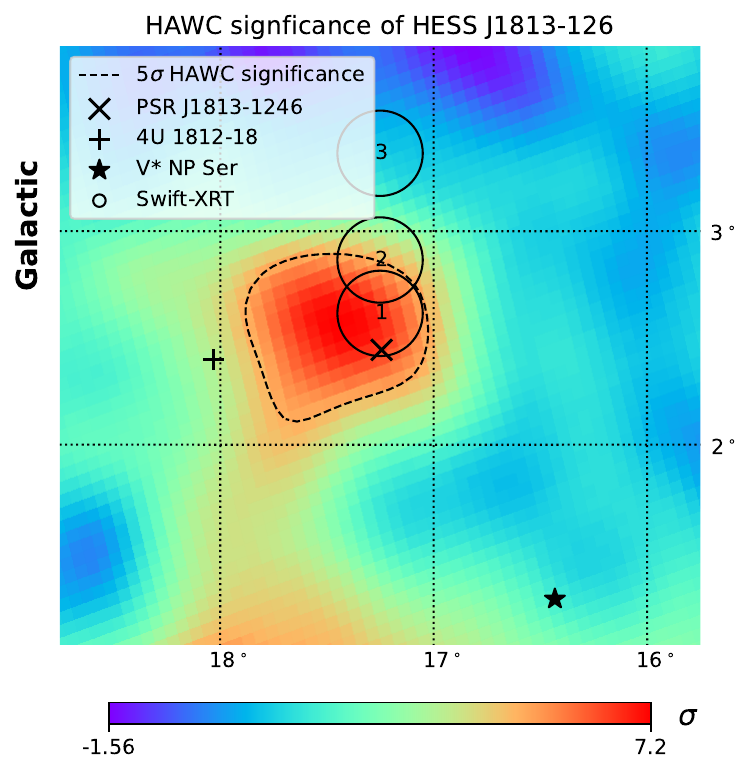}
\caption{The region surrounding HESS J1813-126 is shown above. The color indicates the significance of the HAWC detection with the thick dashed black line indicating the $3\sigma$ level. The HAWC significance maps are produced with public data\footnote{The HAWC public data was retrieved from \url{https://data.hawc-observatory.org/datasets/3hwc-survey/index.php}}. The likelihood is calculated as if there is point source present at the center of each pixel in a HEALPIX pixelation of the sky. The normalization of the source is varied until the likelihood is maximized assuming a spectral index of 2.5. The statistical significance is evaluated as the square root of the log-likelihood ratio \citep{3HWC}. The thin black circles show the \swift field of view centered on the observed targets. 4U 1812-18, PSR J1813-1246, and V* NP Ser are shown as a cross, plus, and star.
\label{fig:hess_skymap}}
\end{figure}

\subsection{Data Processing}

We processed the \swift observations using the standard data processing pipeline following the method of \cite{guevelLimitsLeptonicTeV2023}.
Each field was observed multiple times, so the exposure maps for each individual visit to a field were summed to produce a single exposure map.
The event lists were merged to produce a single event list from which a spectrum will be extracted.
Effective area files were generated using the \texttt{xrtmkarf} with the option for extended sources which corrects for the detector vignetting.
The appropriate response matrices were downloaded from HEASARC \citep{nasahighenergyastrophysicssciencearchiveresearchcenterheasarcHEAsoftUnifiedRelease2014}.

The on-source observations show signs of the bright Earth effect.
This occurs when visible light from the limb of the Earth is scattered on to the XRT detector.
In these observations, the light appears as a brighter than expected region on one side of the detector.
We exclude regions affected by scattered optical light from further analysis.
We produce unfiltered images for each field.
We use \texttt{Ximage} to identify point sources with detection significance greater than $3\sigma$ in these images and exclude a 30\arcsec\, region surrounding each point source.
Visually inspecting each image after point source removal revealed no additional point sources.

After removing detector regions affected by bright Earth light and resolved point sources, we extract a spectrum from each of the three targets.
The spectra are grouped such that each bin contains at least one count.
To account for the different solid angles observed in each of the three targets after excluding these detector regions, we update the \texttt{AREASCAL} keyword in the FITS header in fields 1 and 2.
The value of the \texttt{AREASCAL} keyword in the spectrum for field 1 or 2 is the ratio of pixels included in field 1 or 2 to the number of pixels in field 3.
The values are 0.65 and 0.71.
Any reported surface brightnesses in this paper are corrected by this factor to consistently report the surface brightness across the three fields.

\begin{figure}[t!]
\centering
   \includegraphics[width=0.49\textwidth]{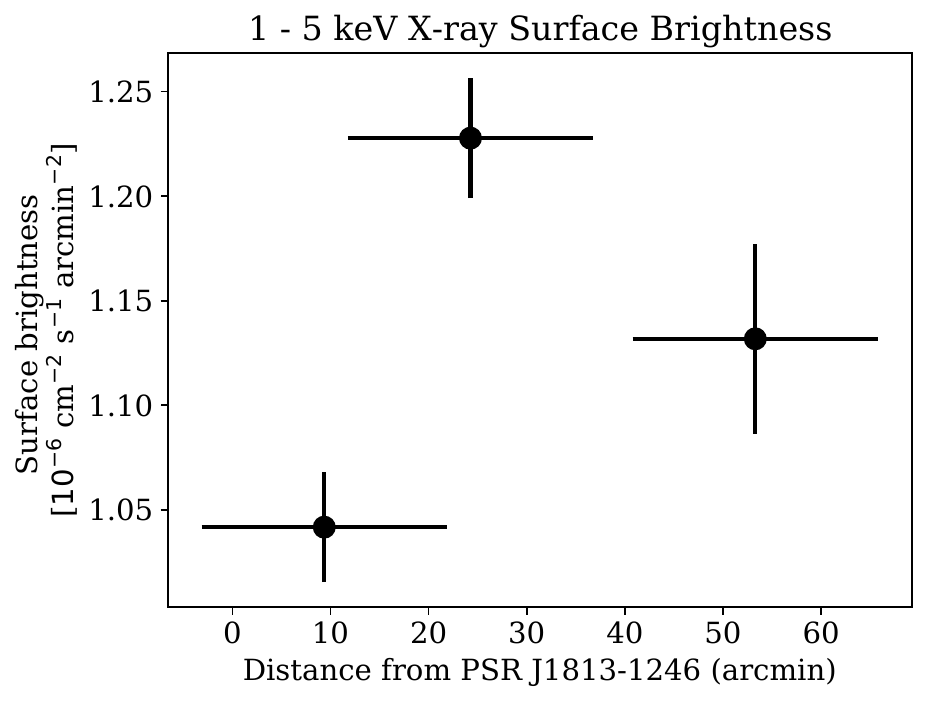}
\caption{The X-ray surface brightness for the three fields is shown as a function of distance from PSR J1813-1246 after excluding point sources and bright Earth contamination. The vertical error bars are the statistical Poisson uncertainty. The horizontal error bars are not statistical; they represent the angular extent of the \swift field of view. The innermost observation slightly overlaps with the pulsar location, so the errors appear beyond zero radius.
\label{fig:radial_profile}}
\end{figure}

\begin{deluxetable}{cccccc}
\tablenum{1}
\tablecaption{Summary of \swift Observations\label{tab:observations}.}

\tablehead{
\colhead{Field} & \colhead{Target ID} & \colhead{l [Deg]} & \colhead{b [Deg]} & \colhead{Exposure [s]}}
\startdata
  1 & 97378 & 17.262923 & 2.599313 & 35.9 \\
  2 & 97379 & 17.249993 & 2.848798 & 32.2 \\
  3 & 97380 & 17.271175 & 3.332234 & 8.7 
\enddata
\end{deluxetable}

\begin{figure}[t!]
\centering
   \includegraphics[width=0.49\textwidth]{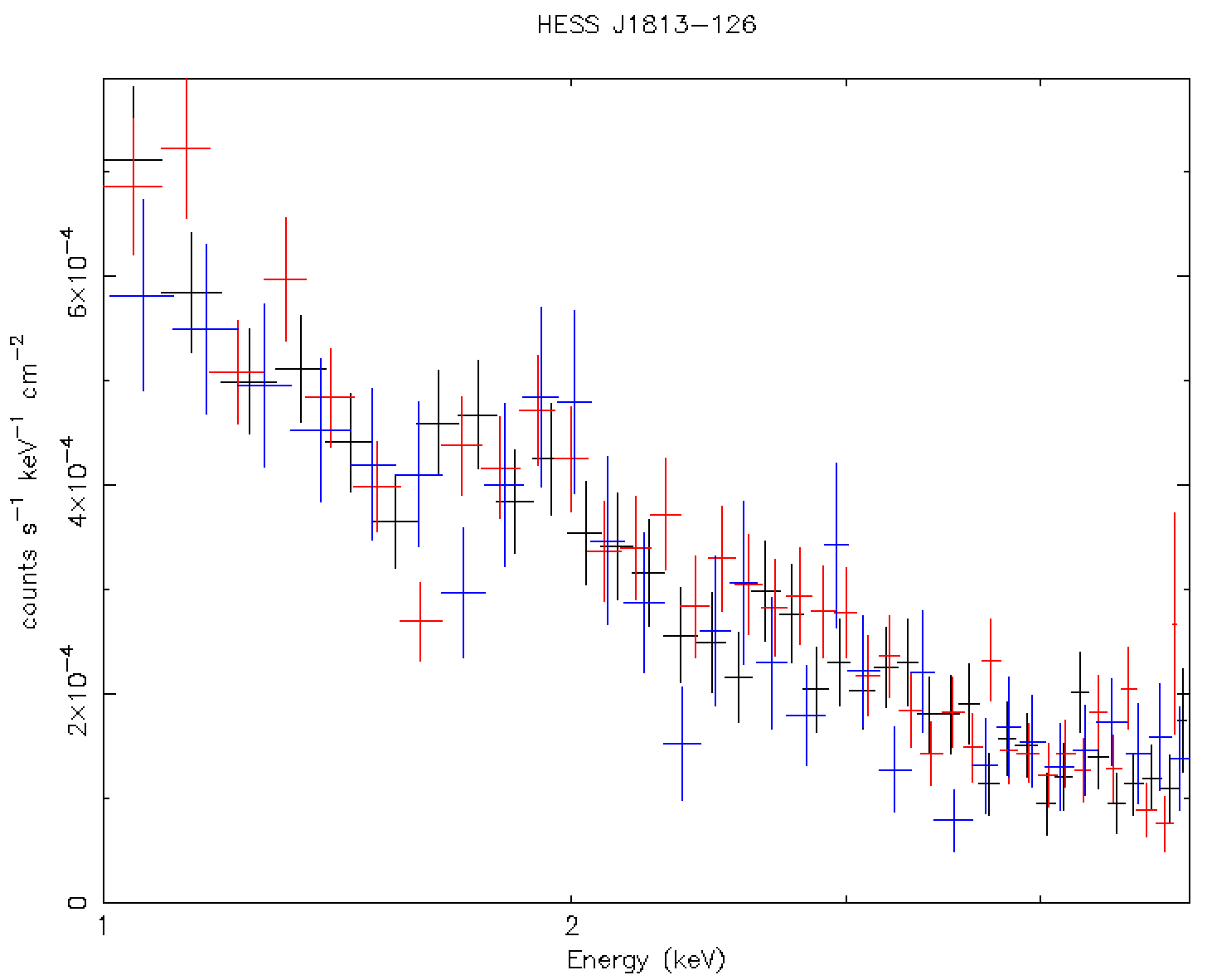}
\caption{Spectra for fields 1 (black), 2 (red), and 3 (blue). The spectra have been produced after removing bright Earth contamination and point sources. The \texttt{AREASCAL} keyword has been adjusted for fields 1 and 2 to account for the sightly different detector area used compared to field 3. The flux of the three spectra are grossly compatible and consistent with GRXB and CXB.
\label{fig:spectra}}
\end{figure}
\begin{deluxetable*}{ccccc|c}
\tablenum{2}
\tablecaption{Summary of Fit Parameters\label{tab:parameters}.}

\tablehead{\colhead{Field} & \colhead{$kT_1$} & \colhead{norm$_1$} & \colhead{$kT_2$} & \colhead{norm$_2$} & \colhead{$\phi$(1 keV)} \\ \colhead{} & \colhead{keV} & \colhead{} & \colhead{keV} & \colhead{$\times 10^{-3}$} & \colhead{$\times 10^{-4}$ keV$^{-1}$ cm$^{-2}$ s$^{-1}$}}
\startdata
  1 & $0.12^{+0.01}_{-0.01}$ & $3.21^{+2.98}_{-1.66}$ & $2.63^{+0.43}_{-0.33}$ & $5.29^{+3.75}_{-3.84}$ & $< 4.32$ \\
  2 & -- & -- & -- & -- & $< 5.38$ \\
  3 & -- & -- & -- & -- & $0^\dagger$ \\
\enddata
\tablecomments{The left side of the table contains the fit parameters in the background only model fit. The dashes indicate that the fit parameter was linked across observations. The right side of the table shows the upper limits on the halo component which is modeled with a power law. For the purpose of measuring the upper limit, the spectral index was frozen to 2. The dagger ($\dagger$) indicates that the parameter was frozen.}
\end{deluxetable*}

\section{Spectral Analysis}
The synchrotron emission from the TeV halo is expected to follow a power law spectral energy distribution in the energy range probed by \swift.
In a one-zone model consisting of a population of electrons propagating in a magnetic field of uniform strength and IC scattering with the cosmic microwave background, the X-ray and $\gamma$-ray emission will have the same spectral energy distribution.
If distribution of electron energies follow a power law, the emission will also follow a power law.
If statistically significant non-thermal emission is observed, more sophisticated models that account for non-uniform magnetic field strength and diffusive energy losses can be tested.
The H.E.S.S. TeV emission is well described by an $E^{-2}$ spectrum, so this is the baseline model that we test for the X-ray emission.
It is unlikely that the TeV halo extends as far as the background field because the electrons will cool before reaching the outskirts of the halo.
The cooler electrons will not emit synchrotron radiation in the X-ray band.
For this reason, we do not search for a signal in the outermost field and use it as an purely background region.
We expect the signal to peak at a few keV where the galactic background is lower, so we fit the spectrum above 1 keV.
Above 5 keV, the spectrum is dominated by the instrument background.

Despite being nearly at the galactic anticenter, there is still significant galactic X-ray emission from $k_T = 0.1 - 3.0$ keV thermal plasma in the galactic disk and non-thermal X-ray emission from distant unresolved AGN.
Above 1 keV, the galactic X-ray emission can be modeled with a two temperature thermal plasma as in \citet{sasakiEROSITAStudiesCarina2024}.
The non-thermal emission, called the cosmic X-ray background (CXB), is modeled with a power law.
Both of these are absorbed by interstellar gas and dust.
Because we only look at emission above 1 keV, we neglect the solar wind charge exchange (SWCX).

The Galactic component is called the Galactic Ridge X-ray Background (GRXB).
It typically requires a two temperature thermal plasma model where the two components are referred to as the cold GRXB and the hot GRXB.
The cold GRXB is emitted by the circumgalactic medium temperature and has a characteristic temperature of $kT\approx0.1$ \citep{pontiAbundanceTemperatureOuter2023}.
The hot GRXB is due to unresolved M dwarf stars and has a typical temperature of a few keV \citep{revnivtsevDiscreteSourcesOrigin2009}.
We model the cold and hot GRXB with two \texttt{APEC} thermal plasma models \citep{smith2001collisional}.

The final X-ray emission component is the cosmic X-ray background (CXB), which is the X-ray emission from the ensemble of unresolved faint AGN.
The CXB is well measured, so we model this component with a power law with spectral index 1.47 and normalization $5.1\times10^{-4}$ keV$^{-1}$ cm$^{-2}$ s$^{-1}$ based on the measurement by \citet{morettiNewMeasurementCosmic2009} and scaling the normalization by the \swift field of view, roughly a circular region with diameter 23.6\arcmin.
The CXB model parameters are not allowed to vary during the fitting.
The GRXB and CXB are subject to absorption by interstellar medium.
We use the \texttt{tbabs} model and freeze the column density to $1.56\times 10^{22}$ cm$^{-2}$ as found by \citet{marelliPuzzlingHighenergyPulsations2014}.
The \swift instrumental background dominates the spectra above 5 keV.
We include the \swift instrumental background as a fixed background spectra.

We search for statistically significant X-ray emission using a likelihood ratio test.
Minimizing the c-statistic is equivalent to maximizing the Poisson log-likelihood \citep{cashParameterEstimationAstronomy1979}, so the likelihood ratio test statistic can be written in terms of the change in c-statistic between the null and test hypothesis, $\textrm{TS}=\Delta C$.
The null hypothesis is that the spectra for the three fields follow the same GRXB plus CXB model.
We simultaneously fit the three fields between 1 and 5 keV with the plasma temperature and normalization parameters linked across all three fields totaling four free parameters.
The test hypothesis is that there is an additional power law component in fields 1 and 2.
The power law index is linked in these two fields but free to vary.
This adds three free parameters
In both these model fits, the c-statistic is minimized for all spectra simultaneously.
In the background only fit, we include the cold and hot GRXB, and CXB in the model.
The null hypothesis fit statistic after minimization is 938.82 with 1056 degrees of freedom.
This suggests that the background only model is consistent with the data.
We added the additional power law component to describe a potential non-thermal component contributed by the halo.
The spectral index is allowed to vary in the fit to accommodate a potential larger range of models.
The normalization of this component is free in fields 1 and 2.
Field 3 is far from the TeV halo, so the normalization of the power law is fixed to zero in field 3.
The spectral index is linked across fields 1 and 2.
The test hypothesis c-statistic is 935.09, so $\textrm{TS}=3.73$ with three additional degrees of freedom.
Under Wilks' theorem, the change in c-statistic should follow a $\chi^2$ distribution with three degrees of freedom \citep{wilksLargeSampleDistributionLikelihood1938}.
This corresponds to a p-value of 0.29 which is not statistically significant.

\begin{figure*}[t!]
\centering
   \includegraphics[width=\textwidth]{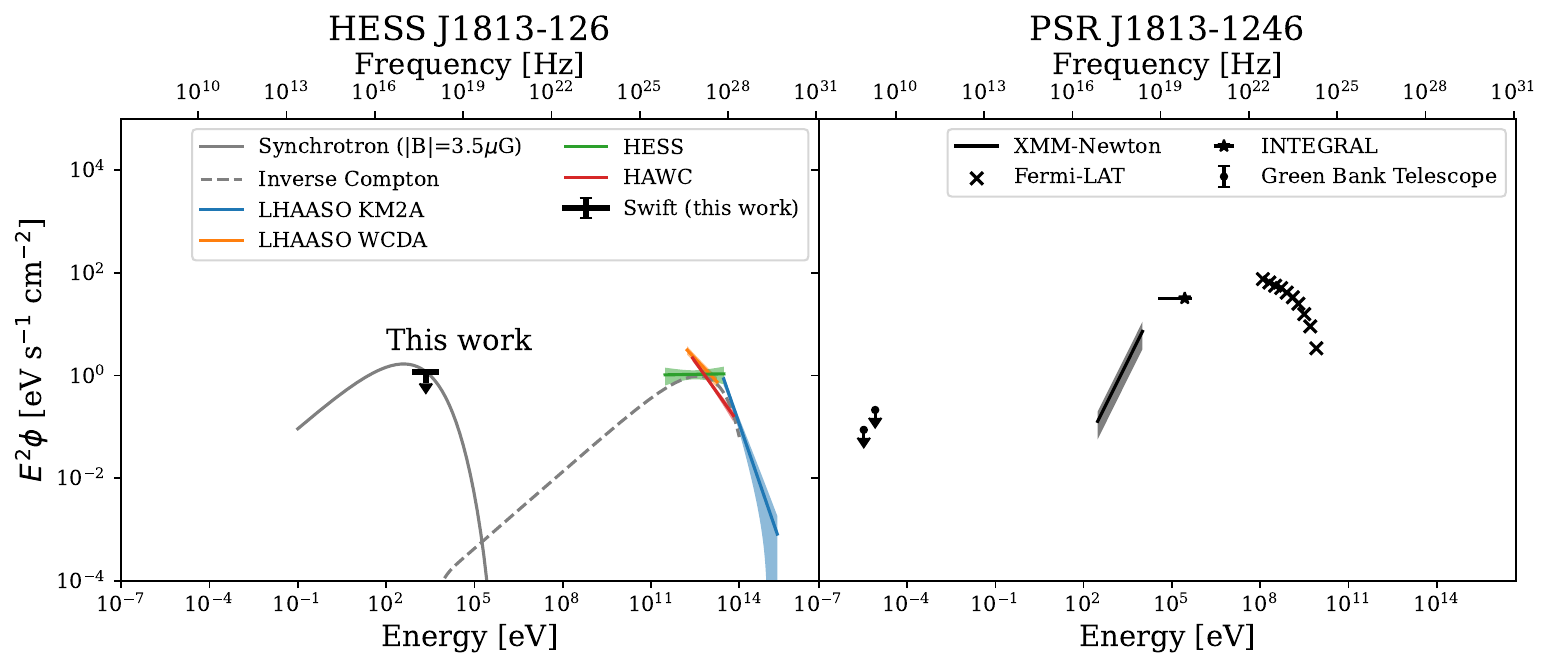}
\caption{The spectral energy distribution for the TeV halo HESS J1813-126 with \swift upper limits (left) and the spectral energy distribution of the pulsar itself (right). The TeV emission is produced by IC scattering with the cosmic microwave background. The synchrotron radiation is produced by the same population of energetic electrons with a magnetic field of uniform strength.
\label{fig:sed}}
\end{figure*}

We place a 90\% confidence level upper limit on the halo's flux normalization by varying the value of the parameter until the c-statistic increases by 2.706 assuming an $E^{-2}$ spectrum.
The limits on the fluxes of such a power law component in field 1 and 2 are obtained as  $4.32 \times 10^{-4}$ keV$^{-1}$ cm$^{-2}$ s$^{-1}$ and $5.38 \times 10^{-4}$  keV$^{-1}$ cm$^{-2}$ s$^{-1}$ at 1 keV, respectively. 

\section{Discussion and Conclusion}

The \swift field of view samples a subregion within the TeV halo.
The observed flux from this subregion cannot describe the complete X-ray emission from the source without assuming a spatial distribution of for the energetic electrons throughout the halo.
We assume that the morphology observed by H.E.S.S. linearly traces the electron population due to the uniformity of the cosmic microwave background within the TeV halo.
We also assume that the magnetic field strength within the halo is uniform.
The synchrotron radiation of a single electron depends on the magnetic field and electron velocity vector.
We assume that the magnetic field configuration is sufficiently turbulent that the X-ray emission is isotropic and unpolarized \citep{blumenthalBremsstrahlungSynchrotronRadiation1970}.
We caveat that the actual X-ray flux of the entire halo could be different from our estimation as the magnetic field could be non-uniform within the halo and the density of electrons may further depend on their energy; however, the lack of observed synchrotron emission precludes the use of a more sophisticated model.
H.E.S.S. found a Gaussian morphology with a standard deviation of $\sigma = 0.21\degr$, so the surface brightness, $I_\gamma$, is 
\begin{equation}
    I_\gamma \propto \frac{1}{2\pi\sigma^2}~\exp\left({-\frac{1}{2}\left(\frac{(\alpha-\mu_\alpha)^2 + (\delta-\mu_\delta)^2}{\sigma^2}\right)}\right)
\end{equation}
where $\alpha$ and $\delta$ are the right ascension and declination, and $\mu_\alpha$ and $\mu_\delta$ are the center position inferred by H.E.S.S.
The two X-ray observations are offset from the center of the H.E.S.S. Gaussian fit, so the fraction of the total X-ray emission from a Gaussian halo captured by \swift must be estimated by numerically integrating the Gaussian morphology over the exposed \swift sky footprint for each observation.
The fraction of X-ray emission captured by \swift is 
\begin{equation}
    \int_\omega d\Omega~ \frac{1}{2\pi\sigma^2}~\exp\left({-\frac{1}{2}\left(\frac{(\alpha-\mu_\alpha)^2 + (\delta-\mu_\delta)^2}{\sigma^2}\right)}\right)
\end{equation}
where $\omega$ is the region exposed to {\it Swift}-XRT.
The integral was evaluated using Monte Carlo sampling.
$10^6$ samples were drawn from the two-dimensional Gaussian, and then the fraction of those samples lying within the \swift field of view was calculated.
The two on-source observations capture 37\% and 11\% of the X-ray emission under this model.
Less than 0.1\% of the flux overlaps with the background region.
Based on these percentages, we estimate the X-ray flux upper limits of the entire halo by scaling individual X-ray upper limits proportionally to the percentage of the X-ray flux captured in each field.
The halo X-ray differential flux limits are $1.16\times 10^{-3}$ keV$^{-1}$ cm$^{-2}$ s$^{-1}$ and $4.73\times 10^{-3}$ keV$^{-1}$ cm$^{-2}$ s$^{-1}$

The X-ray upper limits described above focus on the extended region of the TeV halo.
In this estimate, we assumed a uniform magnetic field and that the electron energy did not depend on distance from the pulsar.
The magnetic field may be elevated closer to the pulsar and the electrons may cool while they diffuse to larger radii.
Both of these would lead to higher synchrotron X-ray emission closer to the pulsar.
\citet{marelliPuzzlingHighenergyPulsations2014} performed a search for a a pulsar wind nebula with a 108 ksec XMM-{\it Newton} and 50 ksec {\it Chandra} exposure and found no evidence for extended X-ray emission in the immediate surroundings of the pulsar.
Both of these observations would have revealed a more compact X-ray morphology if it were present, so we restrict our analysis to the extended halo.

Figure \ref{fig:sed} presents the broadband spectral energy  distributions (SED) of HESS~J1813-126 and its pulsar, along with the derived X-ray upper limit.
The pulsar (right panel) has been observed in radio, X-ray, and GeV $\gamma$-rays.
The radio upper limits were obtained by the Green Bank Observatory after the pulsar was observed by {\it Fermi}-LAT \citep{abdoSECONDFERMILARGE2013a}.
The XMM-{\it Newton} observations of the pulsar were obtained by \citet{marelliPuzzlingHighenergyPulsations2014}, as was the Integral hard X-ray flux.
The very extended halo SED is shown in the left panel.
The X-ray limits obtained in this work are marked in black.
The H.E.S.S., HAWC, LHAASO WCDA, and LHAASO KM2A detections of this source are shown in green, red, orange, and blue \citep{3HWC,HGPS,caoFirstLHAASOCatalog2024a}.
The SED plots indicate that the halo and pulsar emissions originate from different components and are generated by distinct physical processes.

The solid and dashed curves in Figure \ref{fig:sed} show a single-zone SED model of the pulsar halo.
In particular, the electron population is assumed to follow a power law spectrum with an exponential cutoff, $dN/dE_e \propto E_e ^{-2} \exp(-E_e / 50\,{\rm TeV})$.
The electrons produce IC emission by up-scattering the CMB. The electron density is fixed to match the observed TeV flux from HAWC at 7 TeV \citep{3HWC}. 
The same electron population is used to model the synchrotron radiation.
We adjust the magnetic field strength until the synchrotron radiation is the maximum allowed by the X-ray upper limit.
For field 1 (shown in Figure \ref{fig:sed}), the field strength is confined to $B\leq 3.5\mu \rm G$.
For field 2, the maximum allowed magnetic field strength is $6.5\mu \rm G$.

The derived upper limits on the magnetic field strength are consistent with the average Galactic magnetic field strength, which is $\sim 6\,\mu$G locally \citep{2001SSRv...99..243B}.
This suggests that  the magnetic field inside the pulsar halo is not significantly enhanced comparing to the ISM.
Consequently, the confinement of  electrons is more likely related to the structure of the magnetic field in the vicinity of a pulsar and particle transport behaviors in a turbulent or intermittent field.

There have been several efforts to detect X-ray emission from TeV halos.
Our constraints on magnetic field strength are consistent with the magnetic field strengths found by other X-ray observations.
\citet{niuDetectionExtendedXray2025} found evidence for enhanced soft X-ray emission surrounding PSR B0656+14 from observations with eROSITA.
The magnetic field strength near the pulsar is found to be 4-10 $\mu$G and decays radially with profile proportional to $r^{-1}$.
\citet{khokhriakovaSearchingXrayCounterparts2024} also used eROSITA to search for 1\degr\, scale extended emission from five pulsars: Geminga, PSR B0656+14, B0540+23, J0633+0632, and J0631+1036.
They found no significant X-ray emission and placed magnetic field strength limits for each pulsar (1.4, 4.0, 3.1, 2.6, and 2.2 $\mu$G).
\citet{manconiGemingasPulsarHalo2024} use observations from XMM-{\it Newton} and NuSTAR to constrain the X-ray emission from the TeV halo associated with the Geminga pulsar.
Their physically motivated diffusion model is consistent with a magnetic field strength of 2$\mu$G.
The magnetic field strength derived in our analyses is somewhat higher (3.5 and 6.5 $\mu$G) due to the loss of some data from Bright Earth emission and the lower effective area of \swift.
Nonetheless, the non-detection of a degree-scale X-ray counterpart to the TeV halo is consistent with observations and non-detections of other TeV halos.

\vspace{2em}

\begin{acknowledgments}
This work has been supported by the Office of the Vice Chancellor for Research at the University of Wisconsin--Madison with funding from the Wisconsin Alumni Research Foundation. D.G. and K.F. acknowledge support from NASA through the Swift Guest Investigator Program (NNH23ZDA001N). K.F. acknowledges support from the National Science Foundation (PHY-2238916) and the Sloan Research Fellowship. This work was supported by a grant from the Simons Foundation (00001470, KF). K.L.P. acknowledges funding from the UK Space Agency. A.L. acknowledges support from the NASA Fermi Guest Investigator Program (NNH23ZDA001N) and NASA MOSAICS Program (NNH23ZDA001N).
\end{acknowledgments}

\bibliography{main.bib}{}
\bibliographystyle{aasjournal}



\end{document}